\documentclass[11pt]{article}
\usepackage{moriond,epsfig}

\def\Journal#1#2#3#4{{#1} {\bf #2}, #3 (#4)}

\def\NPB{{\em Nucl. Phys.} B}
\def\PLA{{\em Phys. Lett.}  A}

\def\PRD{{\em Phys. Rev.} D}
\def\PR{\em Phys. Rev.}

\def\Nature{\em Nature}

\def\APJL{\em Astrophys. J. Lett.}
\def\CQG{\em Class. Quantum Grav.}
\def\IJMPD{{\em Int. J. Mod. Phys.} D}

\def\dd{{\rm d}}

\def\be{\begin{equation}}
\def\ee{\end{equation}}
\def\bea{\begin{eqnarray}}
\def\eea{\end{eqnarray}}

\newcommand{\cS}{c_{_{\rm s}}}

\newcommand{\epsm}{\epsilon _{_{\rm m}}}
\newcommand{\Hu}{{\cal H}}

\begin{document}
\vspace*{4cm}
\title{Perturbations in the Ekpyrotic Scenarios}

\author{Patrick Peter$^1$, J\'er\^ome Martin$^1$, Nelson
Pinto-Neto$^2$, and Dominik J.~Schwarz$^3$}

\address{$^1$Institut d'Astrophysique de Paris, UPR 341, CNRS, 98
boulevard Arago, 75014 Paris, France\\ $^2$C.~B.~P.~F, Rua Dr.  Xavier
Sigaud 150, Urca 22290-180, Rio de Janeiro, RJ, Brazil\\ $^3$I.~T.~P.,
Technische Universit\"at Wien, Wiedner Hauptstra\ss e 8--10, 1040
Wien, Austria}

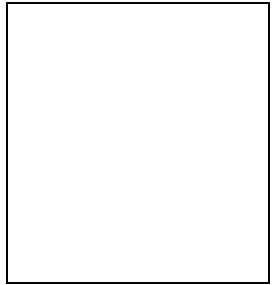
\maketitle\abstracts{With the new cosmological data gathered over the
last few years, the inflationary paradigm has seen its predictions
largely unchallenged. A recent proposal, called the ekpyrotic
scenario, was argued to be a viable competitor as it was claimed that
the spectrum of primordial perturbations it produces is scale
invariant. By investigating closely this scenario, we show that the
corresponding spectrum depends explicitly on an arbitrary function of
wavenumber and is therefore itself arbitrary. It can at will be set
scale invariant. We conclude that the scenario is not predictive at
this stage.}

\section{Introduction}

The standard cosmological model~\cite{standard} is nowadays
well-established on a firm basis consisting of precise observational
data. It supposes in particular a phase of accelerated expansion,
i.e. an inflationary epoch,\cite{inflation} during which the total
energy density of the Universe would have been driven towards its
critical value, the monopole excess diluted and the horizon size
enormously increased. Such a process would thus solve the major
problems of ordinary cosmology, with a bonus: the spectrum of
primordial perturbations it generates, a scale invariant
Harrison-Zel'dovich one, matches the distribution of anisotropies in
the observed micro-wave background sky~\cite{cmb} and the large scale
structure distribution.\cite{lss} With the advent of precision data,
such a prediction happens to be unchallenged as all previously
advocated alternative models have revealed themselves unable to
reproduce these data. The so-called inflationary paradigm, even with
its own difficulties,\cite{transP} thus stands alone on
stage.

By evolving the expanding Universe backwards in time, one can show
that it must have initiated from a singularity.\cite{singularity} This
fact is nowadays assumed to reflect the current limitation of the
understanding of the physics at very high energies at which General
Relativity (GR) in particular does presumably not hold. It was
suggested, back in the thirties,\cite{tolman} by Tolman, Einstein and
Lema\^{\i}tre, that the present phase of expansion might have been
preceded by a collapsing one. Some kind of exotic matter, or violation
of GR was needed for that purpose and the idea sank into oblivion.

Avoiding the primordial singularity through a bounce in a
four-dimensional theory was revived during the
seventies,\cite{seventies} and more recently.\cite{ruth} It was shown
that either some instability develops, in the case of purely
hydrodynamical perturbations,\cite{ppnpn1} or the resulting spectrum
is very much model dependent.\cite{ppnpn2} Another proposal, which is
the subject of this paper, was suggested as a viable competitor to
inflation, and called the ekpyrotic scenario,\cite{ekp,ekpN} includes
a collapsing phase that is extremely slow. Such a slowly contracting
phase was also shown to have the ability to produce a scale invariant
spectrum.\cite{ppnpn2}

In what follows, we briefly recall the basics of the model, discuss
how perturbations are supposed to be produced, and show why they
cannot yet be used in observational cosmology: their calculation
involves an explicit arbitrary function of scale which may be adjusted
in such a way as to reproduce the data.\cite{mpps,comment} The most
difficult part for this model has yet to be done, namely to find a way
of evaluating this arbitrary function on the basis of an underlying
micro-physics.

\section{The ekpyrotic scenario}

The ekpyrotic scenario is supposedly a five-dimensional
superstring~\cite{M} inspired model in which two four-dimensional
branes~\cite{HW} move towards each other, collide, and move
apart. These phases are viewed, in one of the brane, as a slowly
collapsing phase, a singularity which is assumed to provide the
Big-Bang origin of time and to produce entropy, followed by a
subsequent radiation dominated expansion phase. The process may repeat
itself, leading to the so-called cyclic extension.\cite{cyclic} It
should be mentioned that many criticisms have been made to this
proposal, mostly from the viewpoint of particle physics,\cite{pyro}
but also on the applicability of perturbation theory.\cite{Lyth} We
shall not be concerned here with these criticisms and assume that they
have somehow been answered.

From the four-dimensional point of view, it is argued that the model
reduces essentially to GR together with a scalar field $\phi$, the
field evolving in a negative exponential potential, namely
\be {\cal S} = \int \dd^4x \sqrt{-g} \left[ {R\over 2\kappa} -{1\over
2} (\partial \phi)^2-V(\phi)\right], \qquad V(\phi)= -V_{\rm i}
\exp\left[-{4\sqrt{\pi\gamma}\over m_{_{\rm Pl}}}\left( \phi-\phi_{\rm
i}\right)\right],\ee with $\gamma$ a constant and $\kappa=8\pi
G=8\pi/m_{_{\rm Pl}}^2$. The model then yields a
power-law~\cite{powerlaw} collapse up to the point at which the
potential is supposed to vanish. From then on, the scale factor goes
as $a(\eta)\propto\sqrt{|\eta|}$, which vanishes when $\eta$, the
conformal time, changes sign. Later on, radiation takes over, leading
to the standard model with asymptotic behaviour $a\sim\eta$.

Given the four-dimensional model and its matter content thus fixed, we
can in what follows concentrate on the calculation of the spectrum of
primordial perturbations it generates (using notations and conventions
that can be found elsewhere~\cite{mpps}).

\section{Perturbations}

During the slowly contracting phase, labeled in what follows by a
``$<$'' superscript as it refers to negative times, it can be
shown~\cite{mpps,perturbekp,brandenberger} that the Bardeen
gravitational potential acquires a scale invariant piece in the
growing mode. In the long wavelength limit, one gets
explicitly~\cite{mpps}
\begin{equation}
\Phi =B_1(k)\frac{{\cal H}}{a^2} +B_2(k)\frac{{\cal H}}{a^2} \int
^{\eta }\frac{{\rm d}\tau }{\theta ^2}, \qquad B^<_1\propto
k^{-3/2},\qquad B^<_2\propto k^{-1/2},
\end{equation}
with ${\cal H} \equiv a'/a$, $\theta^{-2} \equiv 2\gamma a^2/3$ and
$\gamma=1-\Hu'/\Hu^2$. In the specific case of a free scalar field
dominated collapse, $a=\ell_0 \sqrt{-\eta}/(2\sqrt{2})$ and
$\gamma=3$, this gives a growing mode $\Phi^<_{_{\rm G}}\propto
B^<_1/\eta^2$ and a decaying mode $\Phi^<_{_{\rm D}}\propto B^<_2$.

At the bounce time $\eta=0$, i.e. the singularity, the scale factor
vanishes and the Bardeen potential, in other words the perturbation,
diverges. Even though after that time the linear theory ceases to make
sense,\cite{Lyth} it was nevertheless proposed~\cite{perturbekp} to
use, instead, the comoving density contrast $\epsilon_{\rm
m}=\delta\rho/\rho\propto \Phi/(\rho a^2)$ which is regular at the
bounce. This variable is subject to the differential
equation~\cite{perturbekp} \be\epsm''+ \Hu
\left(1+3\cS^2-6w\right)\epsm'+\left[ 9\Hu^2\left(\cS^2 + {w^2\over 2}
-{4w\over 3}-{1\over 6}\right) + k^2 \cS^2\right] \epsm = - k^2 w
\delta S,\ee where the sound velocity is $\cS^2\equiv\dd p/\dd\rho$,
the equation of state parameter $\omega=p/\rho$, and $\delta S$, the
entropy perturbation, is set to zero for the adiabatic perturbations
here considered. These functions are different on both sides of the
singularity since entropy is produced there. This second order linear
differential equation admits two independent solutions, and its
solution may, in full generality, be written as $\epsm= \epsilon_0
D(\eta) + \epsilon_2 E(\eta)$, with $D(\eta) \sim 1 + 3 \omega_1\eta/4
+ {\cal O}(k^2 \eta^2\ln \eta)$ and $E(\eta) \sim \eta^2 +{\cal
O}(\eta^3)$, where the expansion $w=1+\omega_1\eta+\cdots$ is
assumed. Given the relation between $\Phi$ and $\epsm$, and the
initial value of the Bardeen potential $\Phi^<$, it is a simple matter
to derive the values of $\epsilon_0^<$ and $\epsilon_2^<$ before the
singularity as functions of $B_1^<$ and $B_2^<$.

The quantity that can be compared with observational data is, however,
the Bardeen potential after the singularity, $\Phi^>$, and more
precisely the value of the constant term $B_2$ on the positive time
side of the singularity, written $B_2^>$ (the growing mode of the
collapsing phase then becoming an uninteresting decaying mode in the
expanding phase). This one is given in terms of $\epsilon_0^>$ and
$\epsilon_2^>$, again, after the singularity, and the question arises
then automatically as to what are the matching conditions thanks to
which one may compute these quantities as functions of the pre-bounce
ones?  It has been suggested~\cite{perturbekp} in that regard that the
quantities $\epsilon_0$ and $\epsilon_2$ should keep their numerical
values on both sides of the singularity: $\epsilon_0^>=\epsilon_0^<$
and $\epsilon_2^>=\epsilon_2^<$. With such a choice, the quantity
$B_2^>$ ends up being a linear combination of $B_1^<$ and $B_2^<$,
thereby acquiring a piece proportional to $B_1^<\propto k^{-3/2}$,
i.e. a scale invariant spectrum as announced.

Now it can be argued, as it was indeed pointed out,\cite{perturbekp}
that the normalization of the function $E(\eta)$ must be physically
irrelevant. This amounts to defining a new function $\tilde E = f E$,
and renormalize $\tilde\epsilon_2 = \epsilon_2/f$, with $f$ an
arbitrary function of the background and of the perturbation scale
$k$. Using then the matching condition
$\tilde\epsilon_2^>=\tilde\epsilon_2^<$ then gives the final spectrum
in the form
\begin{equation}
B_2^>(k)=-\frac{9}{8\ell _0^2}B_1^<(k)\left(\frac{f^>}{f^<} \omega
_1^{<2}-\omega _1^{>2}\right) +B_2^{<}(k)\left[\frac{f^>}{f^<}
+2\left(\frac{f^>}{f^<} -\frac{\omega _1^{>2}}{\omega _1^{<2}}
\right)\ln 2\right],\label{result}
\end{equation}
i.e. a spectrum explicitly depending on the arbitrary function $f$ on
both sides of the singularity. It should be mentioned at this point
that setting $\omega_1^<=0$ in this relation does not change the
conclusions since in this case Eq.~(\ref{result}) does not hold and
the calculation must be done differently to yield~\cite{comment} \be
B_2^> = B_2^< {f^>\over f^<} + {9 \omega_1^{>2}\over 8 \ell_0^2}
B_1^<,\ee and the function $f$ has not cancelled out, contrary to what
was claimed in the literature.\cite{perturbekp}

\section{Conclusions}

The inflationary paradigm appears at the present time to be the only
way to explain the shape of the spectrum of primordial
perturbations. The ekpyrotic proposal was suggested as an alternative
that should be confronted with future observations, a task which is
currently impossible since the model predictions depend on an
arbitrary function that can always be made to match the data. As a
result, we conclude that, for the time being and unless some new
physics is specified that explains the matching conditions as well
as sets the arbitrary function, the ekpyrotic scenario is not yet
endowed with any predictive power.

\section*{Acknowledgments}

We would like to acknowledge many illuminating discussions with Robert
Brandenberger, Ruth Durrer, David Lyth, Raymond Schutz, and Gabriele
Veneziano.


\section*{References}
\begin{thebibliography}{99}

\bibitem{standard} E.~W.~Kolb, M.~S.~Turner, {\em The Early Universe},
(Addison-Wesley, Readings, MA, 1990).

\bibitem{inflation}
%A.~Guth, \Journal{\PRD}{23}{347}{1981}; A.~Linde,
%\Journal{\PLB}{108}{389}{1982}; A.~Albrecht and P.~J.~Steinhardt,
%\Journal{\PRL}{48}{1220}{1982}; A.~Linde,
%\Journal{\PLB}{129}{177}{1983}; A.~A.~Starobinsky, {\em Pis'ma
%Zh. Eksp.  Teor. Fiz.} {\bf 30}, 719 (1979) [{\em JETP Lett.} {\bf
%30}, 682 (1979)]; V.~Mukhanov and G.~Chibisov, {\em JETP Lett.} {\bf
%33}, 532 (1981); S.~Hawking, \Journal{\PLB}{115}{295}{1982};
%A.~A.~Starobinsky, {\em Phys. Lett.} B {\bf 117}, 175 (1982);
%J.~M.~Bardeen, P.~J.~Steinhardt, and M.~S.~Turner,
%\Journal{\PRD}{28}{679}{1983}; A.~Guth, S.~Y.~Pi,
%\Journal{\PRL}{49}{1110}{1982}; 
A.~D.~Linde, {\em Particle Physics and Inflationary Cosmology},
(Harwood Academic, New-York, 1990); A.~R.~Liddle, D.~H.~Lyth, {\em
Cosmological Inflation and Large-Scale Structure}, (Cambridge
University Press, Cambridge, UK, 2000).

\bibitem{cmb} C.~Netterfield {\em et al.}, {\tt astro-ph/0104460};
R. Stompor {\em et al.}, {\tt astro-ph/0105062}; P.~De~Bernardis {\em
et al.}, \Journal{\Nature}{404}{955}{2000}; S. Hanany {\em et al.},
\Journal{\APJL}{545}{5}{2000}.

\bibitem{lss} M.~Tegmark, A.~Hamilton, and Y.~Shu, {\tt
astro-ph/0111575}.

\bibitem{transP} J.~Martin and R.~Brandenberger,
\Journal{\PRD}{63}{123501}{2001}; R.~Brandenberger and J.~Martin, {\em
Mod. Phys. Lett.} A {\bf 16}, 999 (2001); J.~Niemeyer,
\Journal{\PRD}{63}{123502}{2001}; M.~Lemoine, M.~Lubo, J.~Martin, and
J.~P.~Uzan, \Journal{\PRD}{65}{023510}{2002}.

\bibitem{singularity} S.~W.~Hawking, G.~F.~R.~Ellis, {\em The large
scale structure of space-time}, Cambridge University Press (1973);
R.~M.~Wald., {\em General Relativity}, Chicago University Press
(1984). See also A.~Borde, A.~Vilenkin, \Journal{\PRD}{56}{717}{1997}
for a more recent discussion.

\bibitem{tolman} R.~C.~Tolman, \Journal{\PR}{38}{1758}{1931};
A.~Einstein, {\em Sitzungsber.}, 235 (1931); G.~Lema\^{\i}tre, {\em
Ann. Soc. Sci. Bruxelles} A {\bf 53}, 5 (1933); R.~C.~Tolman, {\em
Relativity, Thermodynamics, and Cosmology}, (Clarendon Press, Oxford,
1934).

\bibitem{seventies} G. Murphy, \Journal{\PRD}{8}{4231}{1973};
M. Novello and J. M. Salim, \Journal{\PRD}{20}{377}{1979}; V. Melnikov
and S. Orlov, \Journal{\PLA}{70}{263}{1979}; E. Elbaz, M. Novello,
J. M. Salim, and L. A. R. Oliveira, \Journal{\IJMPD}{1}{641}{1993}.

\bibitem{ruth} R.~Durrer and J.~Laukenmann,
\Journal{\CQG}{13}{1069}{1996}.

\bibitem{ppnpn1} P.~Peter and N.~Pinto-Neto,
\Journal{\PRD}{65}{023513}{2002}.

\bibitem{ppnpn2} P.~Peter and N.~Pinto-Neto, {\tt hep-th/0203013}.

\bibitem{ekp} J.~Khoury, B.~A.~Ovrut, P.~J.~Steinhardt and N.~Turok,
\Journal{\PRD}{64}{123522}{2001}.

\bibitem{ekpN} J.~Khoury, B.~A.~Ovrut, N.~Seiberg, P.~J.~Steinhardt
and N.~Turok, {\tt hep-th/0108187}.

\bibitem{mpps} J.~Martin, P.~Peter, N.~Pinto-Neto, and D.~J.~Schwarz,
{\tt hep-th/0112128}

\bibitem{comment} J.~Martin, P.~Peter, N.~Pinto-Neto, and
D.~J.~Schwarz, {\tt hep-th/0204222}.

\bibitem{M} A.~Lukas, B.~A.~Ovrut, and D.~Waldram,
\Journal{\NPB}{532}{43}{1998}~; \Journal{\PRD}{57}{7529}{1998};
A.~Lukas, B.~A.~Ovrut, K.~S.~Stelle, and D.~Waldram,
\Journal{\PRD}{59}{086001}{1999}.

\bibitem{HW} P.~Ho\v{r}ava and E.~Witten,
\Journal{\NPB}{460}{506}{1996}~; {\bf 475}, 94 (1996).

\bibitem{cyclic} P.~J.~Steinhardt and N.~Turok, {\tt hep-th/0111098}.

\bibitem{pyro} R.~Kallosh, L.~Kofman and A.~Linde,
\Journal{\PRD}{64}{123523}{2001}; R.~Kallosh, L.~Kofman, A.~Linde and
A.~Tseytlin, \Journal{\PRD}{64}{123524}{2001}.

\bibitem{Lyth} D.~H.~Lyth, {\tt hep-ph/0106153}; ibid. {\tt
hep-ph/0110007}.

\bibitem{powerlaw} L.~F.~Abbott and M.~B.~Wise,
\Journal{\NPB}{244}{541}{1984}.

\bibitem{perturbekp} J.~Khoury, B.~A.~Ovrut, P.~J.~Steinhardt and
N.~Turok, {\tt hep-th/0109050}.

\bibitem{brandenberger} R.~Brandenberger and F.~Finelli, {\tt
hep-th/0109004}.

\end{thebibliography}
\end{document}